\documentclass[10pt,letterpaper]{article}
\usepackage{mod}
\usepackage{amscd}
\usepackage{amsmath}

\begin{document}

\title{Holographic generation of micro-trap arrays for single atoms}

\author{S. Bergamini, B. Darqui\'e, M. Jones, \\L. Jacubowiez, A. Browaeys, P. Grangier}

\address{Laboratoire Charles Fabry de l'Institut d'Optique, Centre
Universitaire d'Orsay, France}

\begin{abstract}
We have generated  multiple micron-sized optical dipole traps for
neutral atoms using holographic techniques with a programmable
liquid crystal spatial light modulator. The setup allows the
storing of a single atom per trap, and the addressing and
manipulation of individual trapping sites.
\end{abstract}


\section{Introduction}
In the search for a suitable system for quantum information
processing, certain requirements have to be met \cite{04}, such as
 scalability of the physical system,
the capability of initializing and reading out the qubits, and the
possibility of having a set of universal logic gates. Neutral
atoms are one of the most promising candidates for storing and
processing  quantum information. A qubit can be encoded in the
internal or motional state of an atom,  and several qubits can be
entangled using atom-light interactions or atom-atom interactions.
Schemes for quantum gates for neutral atoms have been
theoretically proposed, that rely on dipole-dipole interactions
\cite{02,qg1,qg2,qg3} or controlled collisions
\cite{03,jaksch,10,12}. Such schemes can be implemented in optical
lattices with a controlled filling factor, as shown in ref.
\cite{bloch} where multi-particle entanglement via controlled
collisions was demonstrated.

Presently a major challenge 
is to combine controlled collisions with the
loading and the addressing of individually trapped atoms.
Recently techniques to confine single atoms in
micron-sized \cite{07,RS,01} or larger \cite{meschede}
dipole traps have been experimentally demonstrated.  A set of qubits
can be obtained by creating an array of such dipole traps, each
one storing a single atom \cite{register}.
Gate operations require the
addressability of individual trapping sites and reconfigurability
of the array. An array of dipole traps
can be obtained by focusing a laser beam  into a MOT
with an array of microlenses, as demonstrated in ref. \cite{09}
where each trap could be addressed individually, but where
each trapping spot still contained many atoms.

An alternative method to generate an array of very small dipole traps is
using holographic techniques. Holographic optical tweezers use a
computer designed diffractive optical element to split a single
collimated beam into several beams, which are then focused by a
high numerical aperture lens into an array of tweezers. Recently
holographic optical tweezers have been implemented by using computer-driven
liquid crystal Spatial Light Modulators (SLM)
\cite{grier}. The advantage of these systems
is that the holograms corresponding to various arrays of traps
can be designed, calculated  and optimized on a computer.
Then the traps can be controlled and reconfigured
by writing these holograms on the SLM  in real-time,
and for instance each site can be moved and switched on and off
independantly from the others.

In the present article we present an  experimental
demonstration of the generation of multi-traps arrays for single atoms.
We use a SLM  to control the optical potential of  each trap and
the geometry of the array, and our system allows each single-atom site
to be addressed. This should open an avenue for qubits initialization and readout.

\section{The Spatial Light Modulator (SLM)}

We used the Hamamatsu SLM X7550 spatial light modulator. The SLM
behaves as a mirror which can encode a two-dimensional phase
pattern on the reflected beam, thus acting as a phase grating that
diffracts the light. A prescribed amount of phase shift can be
imposed by varying the local optical path length. This is
accomplished by controlling the local orientation of molecules in
a layer of parallel-aligned (PAL) nematic liquid crystals.
\begin{figure}[h]
\centerline{\scalebox{0.55}{\includegraphics[clip]{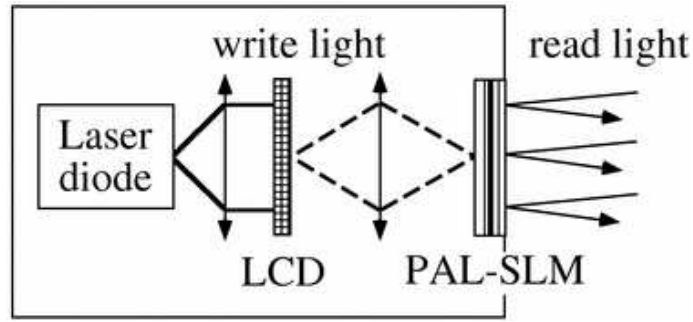}}}
\caption{Scheme of the SLM Module X7550. The surface of the
PAL-SLM is $20 \times 20$ mm$^2$. The LCD is composed of $480
\times 480$ pixels and is controlled by a VGA signal.} \label{slm}
\end{figure}

The structure of the device is described in figure \ref{slm}. The
PAL-SLM module consists of a liquid crystal layer, deposited on a
dielectric mirror. Behind the mirror there is an amorphous silicon
photoconductive layer. This structure is sandwiched between two
transparent electrodes. The orientation of the liquid crystal
molecules is determined by the electric field, which is controlled
locally by changing the impedance of the photoconductive layer
using a write beam, as shown in figure \ref{slm}. The write beam
is intensity modulated using a $480 \times 480$ pixel liquid
crystal device (LCD). Each pixel is controlled individually using
a VGA signal from a computer, and the total active area is $20
\times 20$ mm$^2$. We note that since the readout light is
completely separated from the LCD, diffraction effects due to the
pixellized structure almost vanish, and the optical quality can be
very high.

We measured the optical properties of the PAL-SLM using a Zygo
phase-shift interferometer operating at 633~nm. When the SLM is
switched off the reflectivity is greater than $90\%$, and the
wavefront distortion is 0.6 $\lambda$ peak to peak over the whole
surface  and better than 0.1 $\lambda$ over an active area of side
5 mm. We measured that the phase can be modulated between $0$ and
2.1 $\pi$. For a given optical path length the phase shift is
inversely proportional to the wavelength, so at our operating
wavelength of 810~nm the maximum phase shift
 is reduced to 1.65 $\pi$.

\section{Hologram generation}
The holograms are calculated using an iterative FFT algorithm,
which exploits a  numerical method to calculate the optimal phase
modulation of the incident laser beam in order to obtain  a
desired intensity profile at the imaging plane \cite{05,06}. This
algorithm works in the case of phase-only holograms. We will
consider only regular arrays of optical traps, but the
algorithm may be extended to more complicated structures with no
lattice symmetries.

The basic idea is  to find the relation between the intensity
profile at the focal plane of the focusing objective that we want
to obtain and the necessary phase modulation at the input plane.
The wavefront at the focal plane can be written as~:
\begin{equation}
\label{eq:foc}
E^{f}(\vec{\rho})=E_{0}^{f}(\vec{\rho})\exp[i\phi^{f}(\vec{\rho})]
\end{equation}
and $I^{f}(\vec{\rho})=\vert E^{f}(\vec{\rho}) \vert^2$ is the
intensity profile that we want to obtain. The wavefront at the
entrance pupil of the focusing objective is
\begin{equation}
\label{eq:2}
E^{in}(\vec{r})=E_{0}^{in}(\vec{r})\exp[i\phi^{in}(\vec{r})]
\end{equation}
where $\phi^{in}$ is the phase profile imposed by the  hologram,
that is the pattern we want to calculate. The input wavefront can
be written as the inverse Fourier transform of the wavefront at
the focusing plane:
\begin{equation}
\label{eq:3} E_{0}^{in}(\vec{r})\exp[i\phi^{in}(\vec{r})]=
\mathcal{F}^{-1}\{E^{f}(\vec{\rho})\}
\end{equation}

We start by designing the array of traps we want to obtain in the
focal plane as an  array of Dirac delta functions,
and we obtain  $E_{0}^{f}(\vec{\rho})$ by convoluting the array
with the Airy pattern linked to the entrance pupil of the optical
system. The algorithm is
initialized by a guess of a phase distribution, which is used to
calculate a pattern for the phase modulation of the input wave
$\Phi_{1}^{in}$, as shown in the following diagram. The amplitude $E_{0}^{in}$
is chosen equal to one, as we do not change the amplitude of the
input beam by modulating the phase only.
\begin{equation*}
\begin{CD}
n=1\rightarrow  \Phi_{1}^{in}\\
@VVV\\
 e^{i\Phi_{n}^{in}} @>FFT>>E_{0n}^{f} e^{i\Phi_{n}^{f}}\\
 @A{n\rightarrow n+1}AA  @VVV\\
 E_{n}^{in} e^{i\Phi_{n}^{in}}@<{FFT^{-1}}<< kE_{0}^{f}E_{0n}^{f} e^{i\Phi_{n}^{f}}\\
\end{CD}
\end{equation*}
 Following the diagram above, by Fourier transform we calculate
the image on the focusing plane corresponding to this phase
modulation at the input plane. The result will of course be
different from the desired pattern. At this point the difference
is reduced by multiplying the solution found by the desired
pattern, $E_{0}^{f}(\vec{\rho})$. After normalizing this product
($k$ being a constant to normalize the field amplitude), we take
the inverse Fourier transform of the latter and we extract a phase
pattern for the input beam that is closer to the required one.
Then the cycle is repeated. This kind of algorithm converges
within 3-4 iterations \cite{05,06}.

\begin{figure}[h]
\centerline{\scalebox{0.7}{\includegraphics[clip]{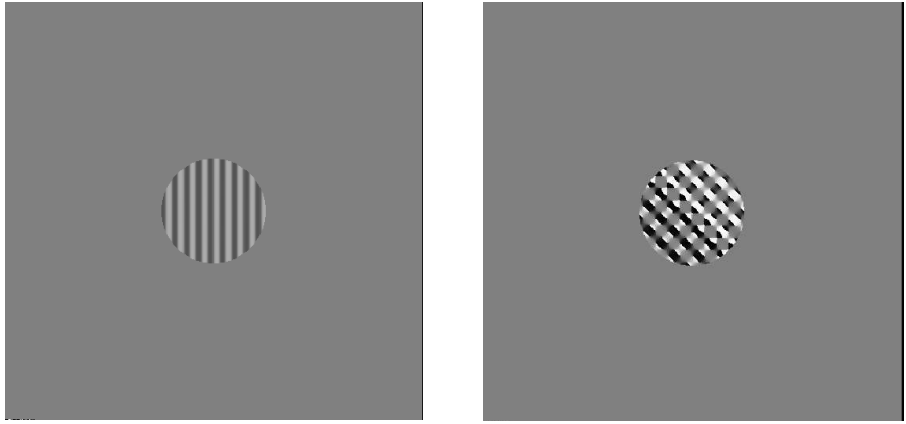}}}
\caption{Examples of two holograms calculated to generate an array
of three dipole traps (left) and five traps in a cross
configuration (right). The different gray levels correspond to
different phase shifts, with black and white giving  a phase shift
$-\pi$ to $+\pi$. For both holograms the separation between the
traps in the focal  plane of the objective is $5$ $\mu$m, and the
pupil size is $5$ mm.}
   \label{holo1}
  \end{figure}

As examples of calculated holograms, figure \ref{holo1} (left)
shows the phase profile of the input wave used in order to obtain
at the focusing plane of the objective an array of 3 dipole traps
in a row separated by $5$ $\mu$m. Figure \ref{holo1} (right) shows
the hologram calculated for obtaining an array of 5 dipole traps.
These holograms were used to generate the trap arrays which will
be presented in the following sections, see figure \ref{diffgeom}.
These patterns are transmitted to the SLM by a standard VGA card.
The different gray levels correspond to different phase shifts,
with black and white giving   a phase shift $-\pi$ to $\pi$.
The phase modulation is calculated taking into account the size of
the beam illuminating the PAL-SLM, so the modulated area in
figure \ref{holo1} corresponds to the size of the beam at the SLM
position. Therefore only a fraction of the total area of the
PAL-SLM is active for phase modulation.

Once the hologram is calculated, it can be modified by changing
several parameters~: the modulation area of the SLM can be reduced,
increased or translated in order to optimize the matching with the
beam's size and position, and the modulation amplitude can be
varied for optimal diffraction efficiency.

\section{The dipole trap experiment}

\begin{figure}[h,b,t]\centerline{\scalebox{0.8}{{\includegraphics[clip]{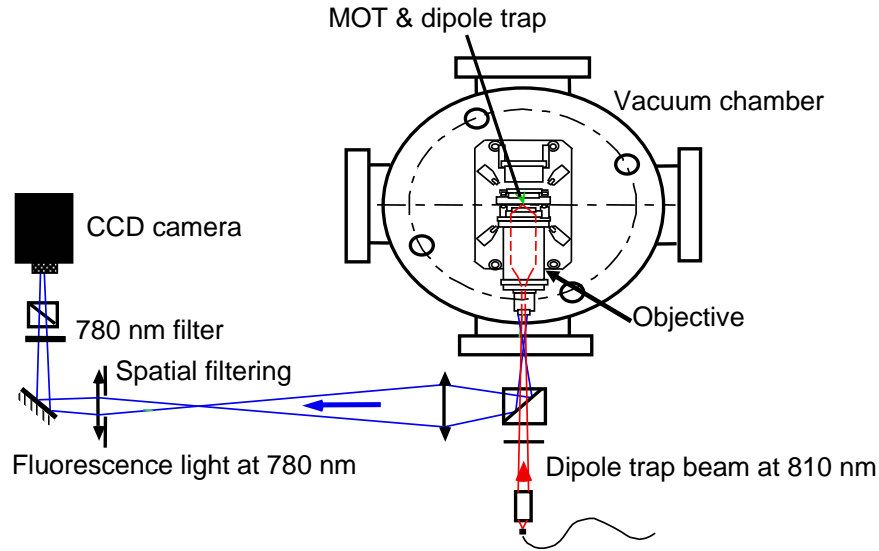}}}}
  \caption{Scheme of the experimental setup (without the SLM device).
The focusing objective (inside the vacuum chamber) generates the
dipole trap at the MOT position. An imaging system collects the
fluorescence light from the trapped atoms and sends it to a CCD
camera}
   \label{chamber}
  \end{figure}

Our apparatus, described in \cite{07,01,RS}, consists of a strongly
focused dipole trap loaded from a magneto-optical trap (MOT) for
Rubidium atoms. The MOT is loaded from an atomic beam, slowed down
by chirped cooling. The dipole trap beam is  focused by an
objective placed inside the vacuum chamber (figure
\ref{chamber}), with a numerical aperture of 0.7. This gives a
measured beam waist of $0.9$ $\mu$m, close to the diffraction limit \cite{RS}.
The effective focal length is $3.55$ mm. This focused beam
provides a tightly confining trapping potential at the center of
the intersection region of the MOT beams. With a relatively small
laser power of $ 10$ mW very high intensities can be reached at
the focusing position ($ 1000$ kW cm$^{-2}$). The dipole trap is
operated in the far-detuned regime, the laser wavelength being
$810$~nm, to be compared to the Rubidium atomic transitions
$D_{1}$ at $795$~nm and  $D_{2}$ at $780$~nm.

The trapped atoms are detected by using the fluorescence induced
by the MOT beams at $780$~nm. The fluorescence is collected by the
same objective which focuses the dipole beam and the detection
system gives a magnified image of the trap on a charge-coupled
device (CCD) camera, as shown in figure \ref{chamber}. A size of
$1$ $\mu$m on the focusing plane of the objective is imaged on 1
pixel of the CCD camera. The integration time is $200$ ms.

The dipole trap  can be operated in several loading regimes
\cite{07,01}. The loading rate of the dipole trap is proportional
to the density of the MOT, which can be varied over several orders
of magnitude by changing the intensity of the magnetic field
gradient and the intensity of the slowing beams. When the MOT
density is small (weak-loading regime) the lifetime of the atoms
in the dipole trap is mainly determined by one-body decay due to
collisions with the background gas. If the loading rate is
increased, due to the very small trapping volume,  there is a
range of loading rates for which two-body collisions become the
dominant term, allowing only one atom at a time to be stored in
the trap. If a second atom enters the trap, a collision occurs and
both atoms are ejected, as shown in ref. \cite{07}. This
``collisional blockade" mechanism operates only for very small
trapping beam waists, typically less than $4$ $\mu$m \cite{07}.
When the MOT density is high  the loading rate is so high that the
average number of atoms in the trap can reach typically $30$
(strong loading regime).
\begin{figure}[h,b,t]
\centerline{\scalebox{0.7}{\includegraphics[clip]{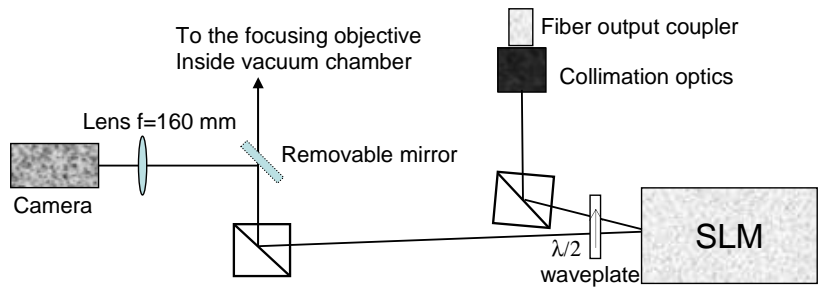}}}
  \caption{Scheme of the experimental setup for phase
modulation of the dipole trap beam. The removable mirror placed in
the beam path is used to send light to an imaging camera, that
records the geometry and shape of the  generated pattern.}
   \label{slmsetup}
  \end{figure}

The dipole trap beam is produced by a $810$ nm laser diode and
brought to the experiment using an optical fiber,
and the PAL-SLM module is placed in the path of the dipole beam,
as shown in figure \ref{slmsetup}. The beam waist at the SLM
position was measured to be $2.3$ mm, so that an area of $\simeq
15$ mm$^2$ of the SLM was illuminated. The power of the incident
beam was varied depending on the number of traps and trap depth
that we wanted to obtain. The SLM can withstand laser intensities
of up to $200$~mW/cm$^2$. In order to maximize the diffraction
efficiency  the incident beam must be linearly polarized along the
0 voltage direction of the molecules, which was ensured by placing
a $\lambda/2$ plate in front of the SLM.

\section{Experimental results}

In this section we show how arrays of traps with different
geometries were created,  by sending  holograms to the SLM. For
simple geometries of the trap array, we can completely extinguish
the trap corresponding to the zeroth-order diffraction spot. We
also prove that we can control the position of the traps with
micrometer precision.  Finally, we used a simple array geometry to
confine single atoms at distinct trapping sites.

\subsection{Tests with different geometries}

Different holograms were calculated with the iterative FFT
algorithm described in section~3. Each calculated hologram was
optimized using an auxiliary lens, with a focal length of 160 mm,
focusing the generated pattern on a standard CCD camera (see figure
\ref{slmsetup}).

\begin{figure}[h]
\centerline{\includegraphics[height=4cm, clip]{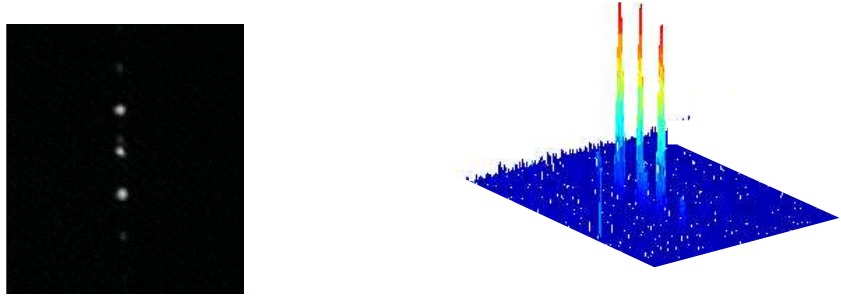}}
  \caption{Two-dimensional and three-dimensional plots of
the intensity profile generated by a three spot hologram.
  The image was captured by focusing the diffracted beams on a CCD camera, by
using an auxiliary lens with  focal length $160$ mm.}
   \label{ccd3-3d}
  \end{figure}

As an example, the intensity profile of a 3-spot array is shown in
figure \ref{ccd3-3d}. The three dimensional plot shows that the
three spots have equal intensity. By adjusting the hologram, as
explained in section 3, we can optimize the symmetry of the
intensity profile, remove higher order diffraction spots, and
control the zeroth-order.
The light was then sent on the atomic sample. We monitored the
resulting fluorescence pattern and fine tuned the hologram in
order to obtain the desired trapping configuration.

\begin{figure}[h]
\centerline{\scalebox{1.1}{{\includegraphics[clip]{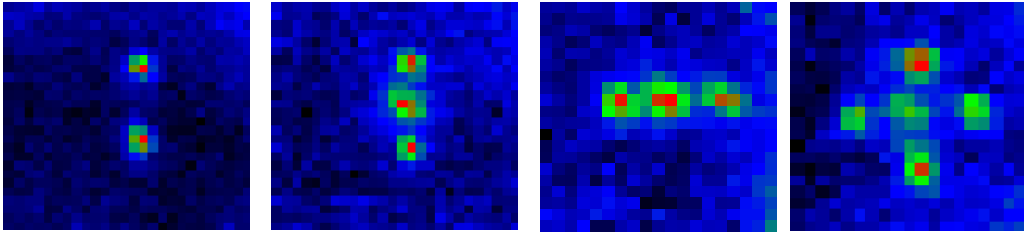}}}}
  \caption{MOT-induced fluorescence of trapped atoms in
dipole trap arrays. The integration time of the CCD is set to
200~ms. The  snapshots show  the different geometries
 tested for a total laser power of 40~mW.}
   \label{diffgeom}
  \end{figure}

The resulting fluorescence pattern from the trapped atoms
for four different holograms is shown on figure \ref{diffgeom}. For these
pictures we worked in the strong loading regime, and so each trap
contains a few tens of atoms. Figure \ref{diffgeom} (left) shows
two traps generated symmetrically with respect to the
zeroth-order, which has been suppressed completely. The second and
third panels of figure \ref{diffgeom} show arrays of three dipole
traps obtained using the phase modulation pattern shown in figure
\ref{holo1} (left), rotated by either $0^{\circ}$  or $90^{\circ}$
in the SLM plane. Finally five traps were generated, using the
hologram of figure \ref{holo1} (right). The limited total laser
power available did not allow us to test structures with larger
number of traps, but we have successfully calculated holograms
with $3 \times 3$ symmetric spots, and hexagonal geometries.
 We limited our tests
to two-dimensional geometries, but three-dimensional
configurations are also possible. The focusing plane could be
changed by adding a lens to the beam path, so that its convergence
can be changed. The effect of the lens can be easily reproduced by
adding a quadratic phase modulation  to the existing hologram.
Therefore, by dynamically changing the computer signal sent to the
SLM it is possible also to create a three-dimensional trap array,
and to change the geometry and the positions of the traps
\cite{08,3d}.

\subsection{Controlling the zeroth order. }

One of the main issues in using an essentially diffractive optical
element is the zeroth-order diffraction spot. Although very high
diffraction efficiency can be achieved, the residual zeroth-order
diffraction spot cannot always be easily controlled for the design
of the pattern. Here we will show that it is possible to either
completely extinguish the trapping site associated to the zeroth
order beam, or to  exploit it to create arrays of equally intense
traps.
\begin{figure}[h]\centerline{\scalebox{0.75}{{\includegraphics[clip]{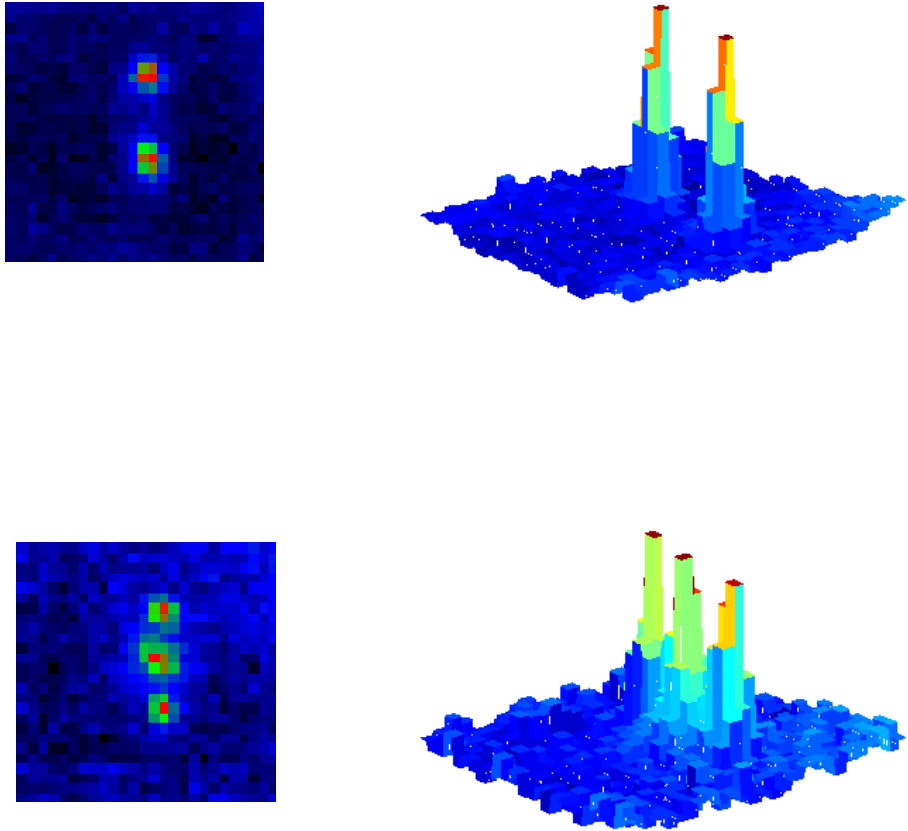}}}}
 \caption{Fluorescence images for two-trap and three trap
arrays, showing control of the zeroth-order. The two traps in the
top figures are generated symmetrically with respect to the
zeroth-order diffraction spot,  while the middle trap in the
bottom is generated from the zeroth-order.}
 \label{2-3-3D}
\end{figure}

In figure \ref{2-3-3D} the images taken for two and three traps
are shown. We note   that the two traps are generated
symmetrically with respect to the zeroth order diffraction spot,
and that the middle trap in the three traps array is generated
with the zeroth-order. We are therefore capable to take into
account the zeroth order in our calculations and control finely
its intensity, so that we can generate arrays of traps where the
zeroth order is suppressed. Actually, we exploit the experimental
observation that there is a trapping depth threshold, below which
the atoms cannot be captured. Therefore, the central peak actually
disappears as soon as the intensity is below the threshold
required to capture atoms, even if the trap light intensity has
not completely vanished. This method provides good enough control
that we are not limited by the  zeroth-order  diffraction spot for
simple geometries.

\subsection{Varying the relative distance}

It is also possible to fully control the relative position
of the trapping sites, either between experiments or dynamically,
just by changing the hologram supplied to the device
as a VGA signal. The lattice constant can
therefore be changed as well as its geometry.

\begin{figure}[h]\centerline{\scalebox{1}{{\includegraphics[clip]{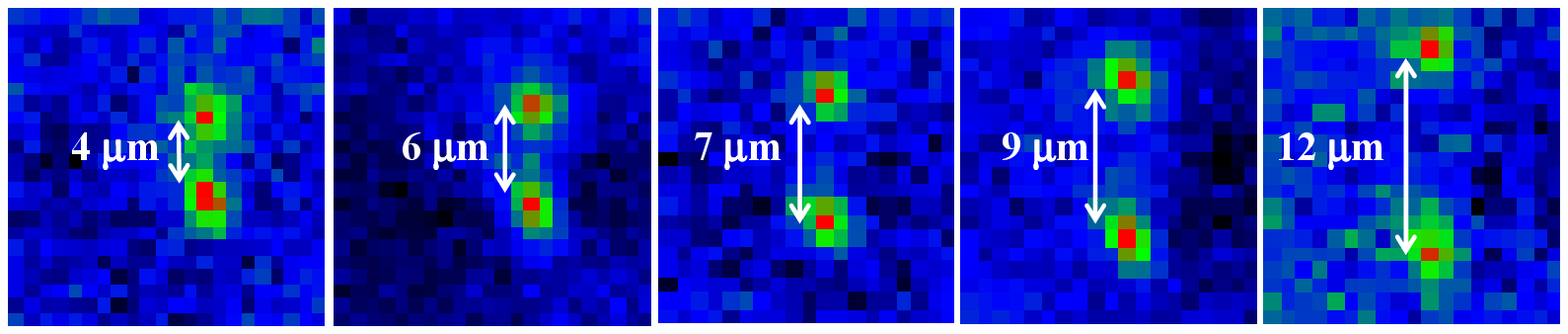}}}}
 \caption{MOT-induced fluorescence of trapped atoms in two
dipole traps.
 The integration time of the CCD is  200~ms. The  figures show
 how the distance could be varied with $\mu$m accuracy  by
 sending a modified signal to the SLM.}
 \label{movie}
\end{figure}
This is illustrated on figure \ref{movie}, where we show how to
control the relative distance of two traps with $\mu$m accuracy.
The trap separation $\delta$ in one array at the imaging plane
depends on the periodicity of the phase modulation:
\begin{equation}
\label{eq:pas} \delta = \frac{\lambda f}{p}
\end{equation}
where $\lambda$ is the laser wavelength, $f$ is the focal distance
of the objective, and $p$ is the phase modulation period. So in
the case of simple geometries, where there is a lattice structure,
the separation between the traps can be changed by modifying the
period of the lattice. Our measurements were limited by the
magnification of the imaging system,  for which one micron on the
focusing plane is imaged on one pixel.

From equation \ref{eq:pas}, the minimum change in the trap
separation is associated to the minimum change in the phase
modulation period, which is given by the size of 1 pixel of the
SLM ($\simeq 40$ $\mu$m). For our present (non-optimized) set-up,
and for the case of $4$ $\mu$m separation of figure \ref{movie},
this gives a limit of precision of the trap position of 300~nm.

In more complicated geometries with many traps, the moving of only
one of the traps with respect to the others can be achieved in
real-time, by sending sequences of pre-calculated holograms to the
SLM. Dynamical control of the trap position is dependent on the
response time of the SLM itself, and on the update rate of the
driving VGA signal. Currently the refresh rate of available
systems using nematic crystals (including the Hamamatsu SLM used
here) is limited to a few tens of Herz. Higher speed (in the kHz
range) can be achieved in principle with commercial ferroelectric
liquid crystals \cite{ferroelectric}, which have however a lower
diffraction efficiency. Therefore the current performance for the
moving speed of the traps does not quite allow fast enough control
for gate operations.

This limitation can
be overcome using schemes which rely on the combination of
an array of (slowly) reconfigurable
traps,  and of a fast ``moving head'', which can be realized with a laser
beam driven by 2D acousto-optic modulators (beam scanners). Such
a scheme would be a neutral atom analog of the proposal of ref.
\cite{zoller} for ion traps.

\subsection{Single atom trapping}

\begin{figure}[h,b,t]
\centerline{\scalebox{1}{{\includegraphics[clip]{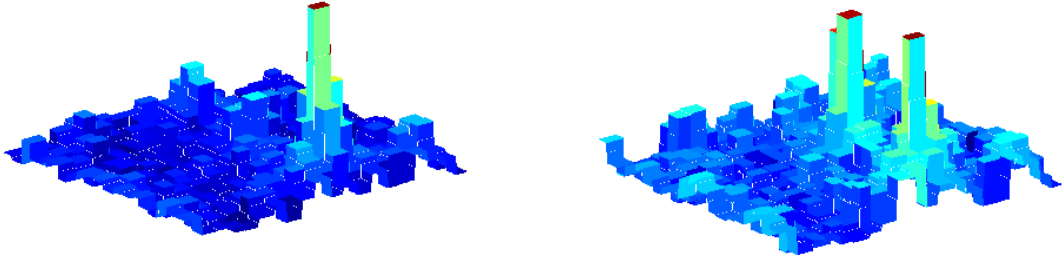}}}}
 \caption{MOT-induced fluorescence of single atoms
confined in distinct dipole traps of a three trap array,
 where the traps are separated by few $\mu$m.
  The two figures show  one atom captured
 in one of the traps and two atoms being simultaneously trapped in two  traps of the three trap array.
 The integration time of the CCD is set to $200$ ms.}
  \label{singleatom}
\end{figure}

Finally, we tested the hologram-generated three-trap configuration
for single atom trapping. By decreasing the density of the atomic
cloud it is possible to enter the regime of loading in which
either one or zero atom is trapped per each site. In figure
\ref{singleatom} (left) a single atom is captured in one trap and
its fluorescence is detected with a 200~ms integration time. On
the right side, two atoms are simultaneously loaded in two
distinct traps. The traps are generated with a laser power of $4$
mW for each one, which is just above the threshold laser power to
capture one atom. Working close to the threshold trap intensity
minimizes the light shift induced by the trapping beam, and
therefore maximizes the MOT-induced fluorescence signal for a
single atom.

Some considerations can be made about the quality of the traps
generated with the hologram. For instance, by comparing the
threshold laser power for traps generated by diffracted beams with
the threshold laser power for trapping with the non diffracted
beam, and assuming that the trapping threshold only depends on the
depth of the trap, we can give a better estimate of the maximum
beam waist enlargement. The trap depth is proportional to the
laser power and inversely proportional to the square of the beam
waist, so if a change in laser power is necessary in order to
reach the trapping threshold this can be easily related to a
change in the beam size. From these considerations we estimate an
upper limit for the waist enlargement of $15$ per cent with
respect to the non diffracted beam, which means an upper limit for
the waist of the diffracted beams  of just over  1 $\mu$m.

In the collisional blockade regime, two-body collisions lock the
average number of atoms to $0.5$ \cite{07}. This means that in
these operating conditions the theoretical probability of
detecting one atom in one of the traps is $0.5$. The probability
of detecting three atoms being simultaneously stored in three
distinct traps therefore  drops down to $0.125$. Unbalance  in the
trap depth would further reduce this probability. Referring to
figure \ref{singleatom}(b),we found that the trap that is not lit
showed a probability $<0.5$ of storing a single atom, which is
probably due to a  shallower trap depth, linked to an asymmetry in
the generated intensity pattern.

\section{Conclusion}

We have demonstrated the possibility of creating multi-trap arrays
for single atoms using a nematic liquid crystal spatial light
modulator. The advantage of using such a device is that it is
fully programmable and computer controllable~: multiple traps can
be generated in different geometries and the position of the traps
can be designed from the VGA signal sent to the module. Arrays of
traps, each capable of storing a single atom, can be dynamically
modified, allowing the real-time motion of one or more traps with
respect to the array.

This  opens up possibilities for testing the proposed schemes for
atom-atom entanglement.
For instance, qubit encoding on the motional state of an
atom in a dipole trap was proposed in \cite{12}. In another
scheme \cite{10}, the qubit is encoded in the motional state of one atom,
which can be trapped in either of two traps with an adjustable separation. Both proposals
\cite{12,10} are studied for Rb micro-dipole traps, for which
single-atom storage has been obtained with our setup. Single qubit
operations are achieved by moving the traps adiabatically and
bringing them closer so that tunneling between the two wells is
allowed, and two-qubit operations are realized via collisions
between two atoms stored in distinct traps.

Alternatively, qubits can be encoded in single atoms
trapped at different locations by using the hyperfine structure of
the ground state, the initialization and single-qubit operations being achieved with Raman
pulses. Two-qubit operations then require either controlled cold collisions
as implemented recently \cite{bloch}, or long-range coupling as proposed in \cite{02,qg1}.

\section*{Aknowledgements}
The authors  acknowledge the contribution of Arnaud Pouderous and Michael
Scholten to early stages of the experiment. We are indebted to S\'ebastien
Bouilhol from Hamamatsu for the loan of the PAL-SLM. This work is supported
by the IST / FET / QIPC project ``QGATES", and by the European
Research Training Network ``QUEST".

\newpage

\end{document}